\documentstyle[12pt]{article}

\input{epsf}
\input{psfig.sty}

\begin{document}

\centerline{{\Large{\bf Relic Gravitational Waves and Cosmology}}\footnote{A 
contribution
to the international conference on cosmology and high-energy astrophysics
``Zeldovich-90" held in Moscow, 20-24 December, 2004; 
http://hea.iki.rssi.ru/Z-90/}}

\vspace{0.8cm}

\centerline{{\large L. P. Grishchuk}}
\centerline{\em grishchuk@astro.cf.ac.uk}

\vspace{0.5cm}

\centerline{School of Physics and Astronomy, Cardiff University, 
United Kingdom}
\centerline{and Sternberg Astronomical Institute, Moscow State University, 
Russia}

\vspace{0.8cm}

\begin{abstract}
The paper begins with a brief recollection of interactions of the 
author with Ya B Zeldovich in the context of the study of relic
gravitational waves. The principles and early results on the 
quantum-mechanical generation of cosmological perturbations are
then summarized. The expected amplitudes of relic gravitational waves are
different in different frequency windows, and therefore the techniques 
and prospects of their detection are distinct. One section of the paper
describes the present state of efforts in direct detection of relic 
gravitational waves. Another section is devoted to indirect 
detection via the anisotropy and polarisation measurements of the 
cosmic microwave background radiation (CMB). It is emphasized  
throughout the paper that the conclusions on the existence and expected 
amount of relic gravitational waves are based on a solid theoretical 
foundation and the best available cosmological observations. I also explain 
in great detail what went wrong with the so-called `inflationary 
gravitational waves', whose amount is predicted by inflationary
theorists to be negligibly small, thus depriving them of any 
observational significance. 
\end{abstract}

\section{Introduction}

The story of relic gravitational waves has revealed the 
character of Ya.~B.~Zeldovich not only as a great scientist but also 
as a great personality. One should remember that the beginning of the 
1970's was dominated by the belief that massless particles, 
such as photons, neutrinos, gravitons, cannot
be created by the gravitational field of a homogeneous isotropic 
universe. Zeldovich shared this view and was publishing papers supporting 
this picture. He was enthusiastic about cosmological particle 
creation \cite{z} and contributed a lot (together with coauthors) to 
this subject. However, he thought that something interesting and 
important could only happen if the early universe was highly anisotropic. 

When I showed \cite{g1, g2} that massless 
gravitons (gravitational waves) could, in fact, be created by the 
gravitational field of a homogeneous isotropic universe, a considerable 
debate arose around this work. I argued that the coupling of gravitons 
to the `external' gravitational field follows unambiguously from the 
equations of general relativity, and it differs from the coupling of other 
known massless particles to gravity. In contrast to other massless fields,
this specific coupling of gravitational waves allows their 
superadiabatic (parametric) amplification by the `pumping' gravitational 
field of a nonstationary universe. (A similar coupling to gravity can 
be postulated for the still hypothetical massless scalar field.) If 
classical gravitational waves were present before the era of 
amplification, they would have been amplified. But their presense 
is not of necessity: even if the waves are initially in their
quantum-mechanical vacuum (ground) state, the state will 
inevitably evolve into a multi-particle state. In phenomenological 
language, gravitational waves are being generated from their zero-point 
quantum oscillations.   

The intense debate has finished in a surprising and very flattering way
for me. It is common knowledge that it was virtually impossible
to win a scientific bet against Zeldovich - he knew practically everything
about physics and had tremendous physical intuition. But sometimes he 
would find a cute way of admitting that his previous thinking was 
not quite right, and that he also learned something from a debate. 
On this occasion it happened in the following manner. 

After one of his rare trips to Eastern Europe (as far as I remember, 
it was Poland) Ya.B. gave me a gift. This was a poster showing a 
sophisticated, impressionist-style, lady. The fact that this was a 
poster with a sophisticated lady was not really surprising - you could 
expect this from Yakov Borisovich. 
What was surprising and flattering for me was his hand-written note at the 
bottom of the poster. In my translation from the Russian, it said 
``Thank you for your goal in my net". Ya.B. was hinting 
at my passion for football, and he knew that this comparison would be 
appreciated much better than any other. So, this is how a great man admits
a clarification of an error; he simply says ``thank you for your goal in 
my net". 

It was clear from the very beginning of the study of relic gravitational
waves that the result of amplification of a wave-field should depend on 
the strength and time evolution of the gravitational pump field. We know 
little about the very early universe these days, even less was known 
at the beginning of the 70's. 

The best thing you can do is to conider plausible 
models. The simplest option is to assume \cite{g1} that the 
cosmological scale factor $a(\eta)$ in the expression
\begin{equation}
\label{metric}
{\rm d}s^2 = a^2({\eta})[-{\rm d}\eta^2 + (\delta_{ij} + h_{ij})
{\rm d}x^i{\rm d}x^j]
\end{equation}
consists of pieces of power-law evolution:
\begin{equation}
\label{a}
a(\eta) = l_o |\eta|^{1+\beta},
\end{equation}
where $l_o$ and $\beta$ are constants. Then, the perturbed Einstein equations 
for $h_{ij}(\eta, \bf{x})$ simplify and can be solved in elementary functions. 
In particular, the intervals of power-law evolution (\ref{a}) make
tractable the effective `potential barrier' $a^{''}/a$ in the 
gravitational wave (g.w.) equation \cite{g1}:  
\begin{equation}
\label{mu}
\mu^{\prime\prime} + \mu \left[n^2 - \frac{a^{\prime\prime}}{a}\right] = 0, 
\end{equation}
where ${}^{\prime} =d/d\eta = (a/c)d/dt$.

Using Eq.(\ref{a}) and the unperturbed Einstein equations one can also 
find the effective equation of state for the `matter', whatever it is, 
which drives the intervals of $a(\eta)$:
\begin{equation}
\label{w}
\frac{p}{\epsilon} = w = \frac{1-\beta}{3(1+\beta)}. 
\end{equation}

The somewhat strange form of the index $1+\beta$ in Eq.(\ref{a}) was
motivated by a serious concern of that time - it was necessary to prove
that even a small deviation from the exceptional law of evolution 
$a(\eta) \propto \eta$ guarantees the effect of g.w.\ amplification.
It is only in this exceptional case that the effective potential 
$a^{''}/a$ vanishes, and therefore the superadiabatic coupling of 
gravitational waves to the nonstationary pump field $a(\eta)$ also
vanishes. 
(The analogous effective potential is absent in equations for photons,
massless neutrinos, and some massless scalar particles.) 

The convenience of the notation utilized in Eq.(\ref{a}) is that 
it parameterises the exceptional case by $\beta = 0$ and 
deviations from this case by a small
$\beta$. Indeed, it was shown \cite{g1}
that the amplitude of the generated g.w.\ mode is proportional to 
small $\beta$; but it is not zero if $\beta \neq 0$. At the same time, 
if $\beta$ is not especially small, the amplitude of the gravitational
wave $h_p(n)$, soon after the beginning of the superadibatic regime and 
while the wave is still in this regime, i.e. before any further 
processing of the amplitude, evaluates to
\begin{equation}
\label{hb}
h_{p}(n) \approx \frac{l_{Pl}}{l_o} \left(\frac{n}{n_H} \right)^{2+\beta}.
\end{equation}

Estimate (\ref{hb}) is approximate (we will be discussing more 
accurate formulas below) but it contains all the necessary physics.
The underlying concepts of generation and detection of primordial
gravitational waves have not changed since the first 
calculations \cite{g1, g2}, and it is important for our further 
discussion to recall them again. 

To begin with, we note that Eq.(\ref{hb}) is formulated for the dimensionless 
amplitude $h$ of a given g.w.\ mode characterised by a constant 
dimensionless wavenumber $n$. (The $h(\eta)$ and $\mu(\eta)$ mode-functions
are related by $h=\mu/a$.) The wavelength $\lambda$, measured in 
units of laboratory standards (as Zeldovich used to say, measured in 
centimeters), is related to $n$ by $\lambda(\eta) = 
2\pi a(\eta)/n$. It is convenient to use (and we will always do this) 
such an $\eta$-parameterisation of $a(\eta)$ that the present-day scale 
factor is $a(\eta_R) = 2 l_H$, where $l_H = c/H(\eta_R)$ is the 
present-day value of the Hubble radius. Then, $n_H=4\pi$ is the 
wavenumber of the waves whose wavelength today is equal the prsent-day 
Hubble radius. Longer waves have smaller $n$'s, and shorter waves have 
larger $n$'s. 

Expression (\ref{hb}) is essentially a
consequence of the following two assumptions.
First, it is assumed that the mode under consideration 
has entered the superadibatic regime in the past, and is still in 
this regime. This means that the mode's
frequency, instead of being much larger than the characteristic 
frequency of the pump field, became comparable with it at some time
in the past. Or, in cosmological context, the wavelength 
$\lambda(\eta)$ of the mode $n$, instead of being much shorter than the 
instanteneous Hubble radius $c/H(\eta) = a^2/a'$, became equal 
to it at some moment of time $\eta_i$, i.e. $\lambda_i =c/H_i$. 
For the scale factors of Eq.(\ref{a}), this condition leads to 
$(n/n_H)|\eta_i| \approx 1$. 

Second, we assume that by the beginning of the superadiabatic regime
at $\eta=\eta_i$, the mode has still been in its vacuum state, rather
than, say, in a strongly excited (multi-particle) state. That is, 
in the language of classical physics, the mode's amplitude near 
$\eta_i$ was not much larger than $h_i(n) \approx l_{Pl}/\lambda_i$, 
where $l_{Pl}$ is the Planck length, $l_{Pl} =\sqrt{G \hbar/c^3}$. 
This condition on the amplitude follows from the 
requirement that initially there were only the
zero-point quantum oscillations of the g.w.\ field, and the 
initial energy of the mode was $(1/2) \hbar \omega_i$. Because 
of the condition $\lambda_i =c/H_i$, we can also write $h_i(n)$ as  
$h_i(n) \approx H_i~l_{Pl}/c$. 

The amplitude of the mode, after the 
mode's entrance to the amplifying superadiabatic regime, and as 
long as this regime lasts, remains at the constant level $h_i(n)$,
i.e. $h_p(n) \approx h_i(n)$. This holds true instead of the adiabatic 
decrease of the amplitude $\propto 1/a(\eta)$ that 
would be true in the adiabatic regime. In general, the quantity $H_i$ is
different for different $n$'s: 
\[
H_i \approx \frac{c}{l_o}  \eta_i^{~-(2+\beta)}
\approx \frac{c}{l_o} \left( \frac{n}{n_H} \right)^{~2+\beta}. 
\]
Therefore, a specific dependence 
on $n$ arises in the function $h_i(n)$, and this is how
one arrives at Eq.(\ref{hb}) in a simple qualitative manner.

Formula (\ref{hb}) gives the evaluation of the primordial (before
further processing) g.w.\ spectrum $h_p(n)$. Roughly speaking, the initial 
vacuum spectrum $h_{v}(n) \propto n$ has been transformed into
the primordial spectrum $h_p(n) \sim h_v(n) n^{1+\beta_i}$, where 
$\beta_i$ characterizes
the scale factor of the era when the transition from the adiabatic to
superadiabatic regime has taken place for the given interval of
wavenumbers $n$. However, the same
mode $n$ can sooner or later leave the amplifying regime and start
oscillating again. Obviously, this reverse transition from superadiabatic
to adiabatic regime is being described
by the same theory. 

The final amplitudes at some fixed moment
of time (for example, today's amplitudes) $h_f(n)$ are related 
to the $h_p(n)$-amplitudes by 
\[
h_f(n) \sim h_p(n) n^{-(1+\beta_f)}, 
\]
where $\beta_f$ characterizes the era when the opposite 
transition from the superadiabatic to adiabatic 
regime has taken place (this is why the minus sign
arises in front of $1+ \beta_f$ in the exponent).

The discussed amplitudes $h(n)$ are in fact the 
root-mean-square ($rms$) amplitudes of the multi-mode field, 
they determine the mean-square 
value of the wave field $h$ according to the general formula
\[
\langle h^2\rangle = \int h_{rms}^{2}(n) \frac{dn}{n}.
\]

It is necessary to say that in the beginning of the 80's, the inflationary 
cosmological scenario governed by a scalar field \cite{guth} was gaining 
popularity. 
Its central element is the interval of deSitter expansion, which 
corresponds to $\beta = -2$ in Eq.(\ref{a}) ($\eta$ grows from
$- \infty$, $1+\beta < 0$) and $w=-1$ in Eq.(\ref{w}). By the time
of publication of the inflationary scenario, 
unusual equations of state for `matter' driving the very early Universe, 
including such exotic ones as $p= -\epsilon,~ w=-1$, had already 
been the subject of cosmological research, most notably in the 
work of A. D. Sakharov \cite{sah}. 

The g.w.\ calculations for 
the special case $\beta = -2$ were performed in a 
number of papers (see for example \cite{star, rsv, 
fp, aw}). If $\beta = -2$, the dependence on $n$ vanishes in the 
general Eq.(\ref{hb}), and the primodial (unprocessed) spectrum 
$h_p(n)$ becomes `flat' -- that is, $n$-independent. Ironically, the 
prospects of direct detection of the stochastic g.w.\ background
characterised by the corresponding processed (today's) spectrum
had already been explored by that time \cite{g2}; the processed spectral 
index for this model is $\alpha = 1$ in notations of that paper.
Ref.\cite{g2} also suggested the use of cross-correlated data from two
detectors and touched upon the technique 
of `drag-free satellites' that was later developed in the Laser 
Interferometer Space Antenna (LISA).

The generality of inflationary, quasi-deSitter, solutions was a serious 
concern for Zeldovich during long time. He kept wondering about the 
sensitivity of inflationary solutions to the choice of initial conditions. 
Nobody would take the inflationary scenario seriously if it were a 
very contrived
or unstable solution. However, it was shown \cite{bgkz} that the 
inflationary-type evolutions are, in fact, attractors in the space of 
all possible solutions of the corresponding dynamical system. This 
decisive property made inflationary evolutions more plausible and 
appealing.

\section{Direct detection of relic gravitational waves}

The spectrum of $h_{rms}(\nu)$ expected today is depicted in Fig.1
(for more details, see \cite{g3, g4}). Almost everything in this 
graph is the result of the processing of the primordial spectrum during 
the matter-dominated and radiation-dominated stages. The postulated
`Zeldovich's epoch' governed by a very stiff effective equation 
of state, is also present in the graph, as shown
by some relative increase of power at very high frequencies. The 
primordial part of the spectrum survives only at frequencies below the 
present-day Hubble frequency $\nu_H \approx 2 \times 10^{-18}$ Hz. 
The available CMB observations 
determine the amplitude and spectral slope of the g.w.\ spectrum at 
frequencies around $\nu_H$, and this defines the spectrum at higher 
frequencies. 

\begin{figure}[!hbt] 
\caption{Envelope of the $h_{rms}(\nu)$ spectrum for the 
case $\beta =-1.9~({\rm n}=1.2)$} 
\centerline{\psfig{file=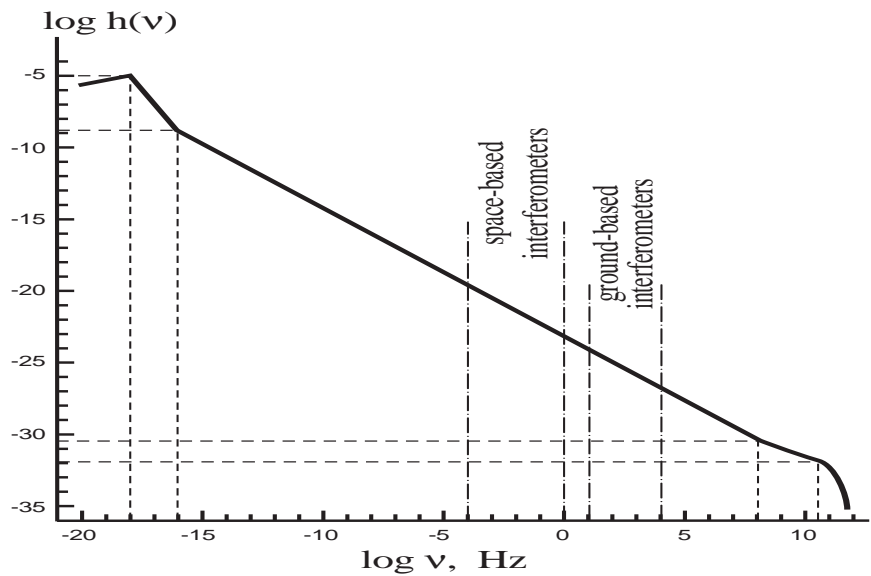,height=3.in,width=4.5in}}
\end{figure}

The numerical value of $h_{rms}$ at frequencies around $\nu_H$ is determined
by the numerical value of the observed quadrupole anisotropy of CMB. As
will be shown in great detail in Sec.4, it follows from the theory
of cosmological perturbations that relic gravitational waves should 
provide a significant fraction of the observed CMB signal at very large 
angular scales (barring the logical possibility that the observed 
anisotropies have nothing to do at all with cosmological perturbations 
of quantum-mechanical origin).

In other words, the final theoretical results do not 
contain any dimensionless parameter 
which could be regulated in such a manner as to make the contribution 
of, say, density perturbations to the quadrupole anisotropy several orders of 
magnitude larger than the contribution of gravitational waves. These 
contributions must be roughly equal, but the 
theory cannot exclude that one of them will turn out to be a numerical 
factor 2-3 larger than another. Assuming that relic gravitational 
waves provide a half of the signal, one can find from the observed
$\delta T/T \approx 10^{-5}$ that $h_{rms}(\nu_H) \approx 10^{-5}$ and,
hence, it follows from Eq.(\ref{hb}) that $l_{Pl}/l_o \approx 10^{-5}$.

The slope of the primordial g.w.\ spectrum is also taken from CMB 
observations. The commonly used spectral index ${\rm n}$ 
(we denote it by a Roman letter ${\rm n}$ in order to distinguish from
the wavenumber $n$) is 
related to the parameter $\beta$ appearing in Eq.(\ref{hb}) 
by the relationship ${\rm n}= 2 \beta +5$. The same relationship 
is valid for density perturbations, to be discussed later.
The current observations \cite{bennet, page} give evidence for  
${\rm n} \approx 1$, which corresponds to $\beta \approx -2$. The 
particular graph in Fig. 1 is plotted for $\beta=-1.9,~ {\rm n}=1.2$, 
which tallies with the COBE data \cite{smoot, gorski}. 
(This spectral index $n > 1$ implies that $w < -1$, according to Eq.(\ref{w}). 
It is not difficult to imagine that such an effective equation of state
could hold in the very early Universe, if the recent supernovae 
observations hint at the validity of $w < -1$ even in the present-day 
Universe !) In simple words, 
the position and orientation of the entire piece-wise function 
$h(\nu)$ is defined by the known value of the function at the 
point $\nu = \nu_H$ and the known slope of the function in the 
vicinity of that point.  

Incidentally, the initial quantum vacuum conditions for gravitational
waves, at all frequencies shown in the graph, are formulated at the 
`initial' moments of time, when each wavelength of interest was 
appreciably longer than the Planck length. Therefore, the shown 
results are immune to the short scale ambiguities of the 
so-called `trans-Planckian' physics (see for example 
\cite{greene}). It is a different matter 
that at some frequencies the initial state is allowed to be a 
somewhat excited state, rather than the pure vacuum state, 
without running into a conflict with the adopted approximation of small
perturbations. This exotic possibility and the corresponding modifications
of the spectrum were discussed long time ago \cite{gs} (see also a 
related work \cite{crei}).
\begin{figure}[!hbt] 
\caption{S3 LIGO noise curves and the expected sensitivity 
$\Omega_0 \sim 10^{-4}$ to stochastic gravitational waves} 
\centerline{\psfig{file=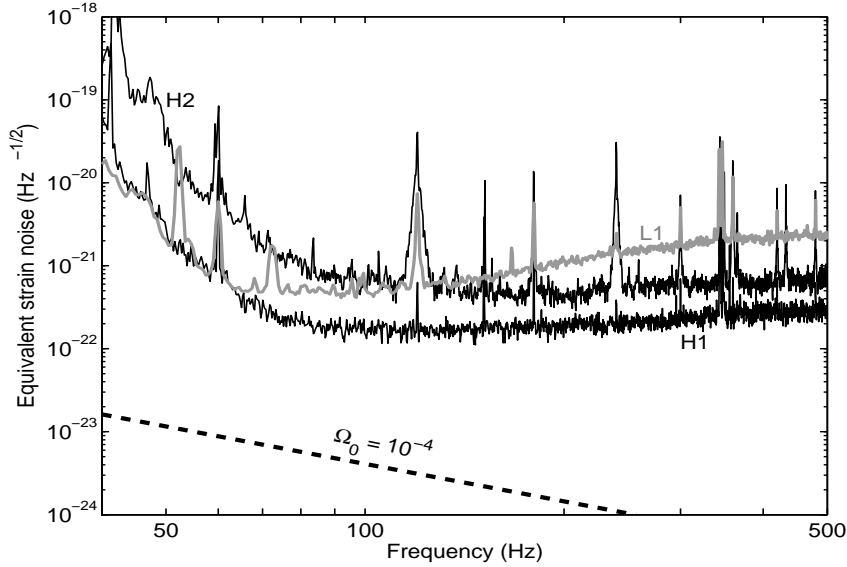,height=3.in,width=4.5in}}
\end{figure}

The graph in Fig.1 shows the piece-wise envelope of the today's spectrum.
The displayed result is quite approximate. In particular, it completely 
ignores the inevitable oscillations in the spectrum, whose origin 
goes back to the gradual diminishing (squeezing) of quantum-mechanical 
uncertainties in the phases of the emerging waves and the macroscopic 
manifestation of this effect in the form of the standing-wave pattern 
of the generated field. (This is also related to the concept of `particle 
pair creation'.) We will discuss these spectral oscillations below.

Nevertheless, the graph in Fig.1 is convenient in that it gives 
simple answers to the 
most general questions on the amplitudes and spectral slopes of relic 
gravitational waves in various frequency intervals. For example, it shows 
the expected amplitude $h_{rms} = 10^{-25}$ at $\nu = 10^2$ Hz. This 
is the level of the signal that we shall be dealing with in experimental 
programs. In terms of the parameter $\Omega_{gw}(\nu)$,
\[
\Omega_{gw}(\nu) = \frac{\pi ^2}{3} h^2(\nu) \left(\frac{\nu}{\nu_H}\right)^2,
\]
it corresponds to $\Omega_{gw} \approx  10^{-10}$ at frequency 
$\nu = 10^2$ Hz and in its vicinity.

Where do we stand now in the attempt of direct detection of relic 
gravitational waves? The sensitivity of the presently operating
ground-based interferometers is not good enough to reach the predicted
level, but the experimenters are making a lot of progress. 
The data from the recently completed S3 run of LIGO \cite{ligo} will 
probably allow one to reach the astrophysically interesting level 
of $\Omega_{gw} \sim 10^{-4}$, as shown in Fig. 2 (courtesy of
J. Romano and the stochastic backgrounds group of LSC). Fortunately, 
the projected
sensitivity of the  advanced LIGO ($\sim$2011) will be sufficient to reach 
the required level of $h_{rms} \approx 10^{-25},~~
\Omega_{gw} \approx 10^{-10}$, when a month-long stretch of 
cross-correlated data from the two detectors is available.

\begin{figure}[!hbt] 
\caption{Various LISA sources including relic gravitational waves} 
\centerline{\psfig{file=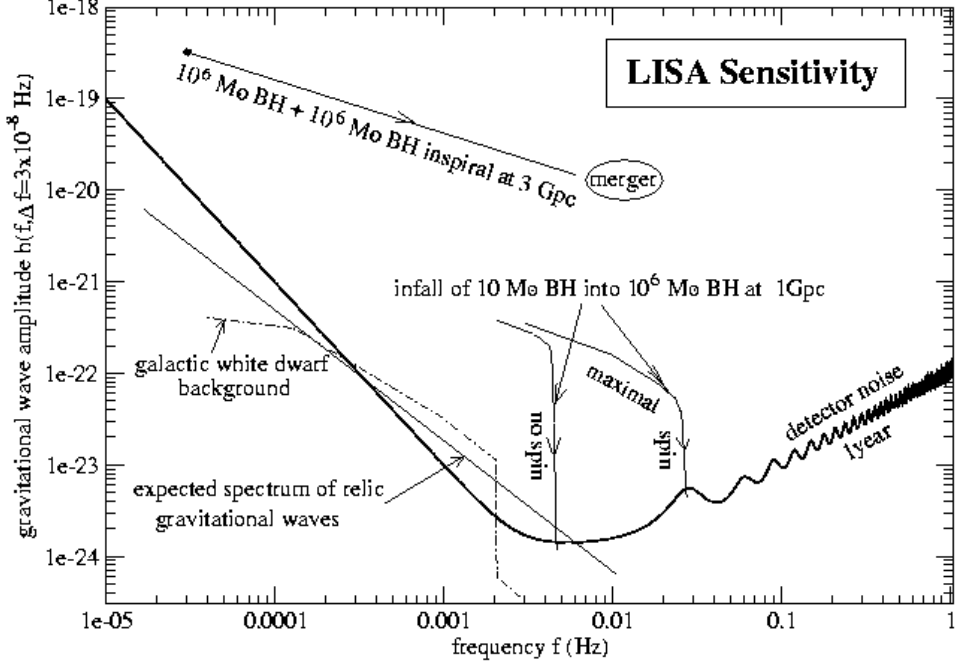,height=3.5in,width=5in}}
\end{figure}

The ESA-NASA space-based mission LISA ($\sim$2013) will have a 
better chance to discover relic gravitational waves. Since the 
expected spectrum has larger amplitudes at
lower frequencies, the detectability condition is potentially 
improving at lower frequencies. In Fig. 3 we show the LISA 
sensitivity in frequency bins of $\Delta f = 3 \times 10^{-8}$ Hz,
which corresponds to an observation time of 1 year. This observation
time should make it possible to resolve the g.w.\ lines from thousands 
of white dwarf binaries in our Galaxy, radiating at frequencies larger 
than $2 \times 10^{-3}$ Hz. By removing the contribution of the binaries 
from the observed records, or by using sophisticated data analysis techniques 
without actually removing the contaminating signals from the data, one can 
effectively clean up the window of instrumental sensitivity at frequencies 
above $2 \times 10^{-3}$ Hz. This window in the area of maximal sensitivity
of LISA is shown in the graph together with the expected level 
of relic gravitational waves in that window.

\section{Indirect detection of relic gravitational waves via 
CMB anisotropies and polarisation} 

The expected amplitudes of relic gravitational waves reach their 
highest level in the frequency interval of $10^{-18} - 10^{-16}$ Hz. 
This is why one has very good prospects for indirect detection of 
relic gravitational waves through the measurements of anisotropies 
in the distribution over the sky of the CMB temperature and polarisation.
(For an introduction to the theoretical tools of
CMB physics, see for example \cite{giovan}.) 

The accurately calculated power spectrum $h^2_{rms}(n)$ 
is shown in Fig. 5 \cite{bgp}. The spectrum is calculated at the moment
of decoupling (recombination) of the CMB, with the redshift of
decoupling at $z_{dec} = 1100$. The derivation of the spectrum
takes into account the quantum-mechanical squeezing of the waves'
phases, which manifests itself macroscopically in the standing-wave
character of the generated gravitational waves. From the viewpoint
of the underlying physics,
it is this inevitable quantum-mechanical squeezing that is
responsible for the oscillations in the power spectrum. 

The displayed spectrum was obtained under the assumption that 
$\beta= -2$ $({\rm n} =1)$, i.e. for a flat 
primordial spectrum. The survived part of the primordial flat spectrum 
is seen on the graph as a horizontal part of the curve in 
the region of very small wavenumbers $n$. The normalisation of 
the spectrum is chosen in such a way that the induced 
quadrupole anisotropy of the CMB today is at the level of the actually 
observed quadrupole \cite{smoot, bennet}. Specifically, the temperature 
function $l(l+1)C_l$ in Fig. 4, calculated from the spectrum 
in Fig. 5, gives the required value of 960 $(\mu K)^2$ at $l=2$. 
The distribution of other induced multipoles is also shown in Fig. 4.

Figures 4 and 5 are placed one under another on purpose. 
This placement gives a better visual description of the fact noticed 
and explained previously \cite{bg}. Namely, the oscillations
in the metric (gravitational field) power spectrum are entirely 
responsible for the
oscillations in the angular power spectrum of the CMB temperature,
with almost universal correspondence between extrema in the wavenumber 
space $n$ and extrema in the multipole moment space $l$. If there 
is much/little power in the gravitational field perturbations of
a given interval of wavelengths, one should expect much/little 
power in the temperature fluctuations at the corresponding
angular scale. 

It is the oscillations in the metric power 
spectrum that are responsible for the oscillations in the 
$l$-space, and not the mysterious explanations often repeated 
in the literature, which claim that the peaks in the function
$l(l+1)C_l$ arise because of some waves being caught (at the moment of 
decoupling) in their maxima or minima, while others are not. 
To illustrate the role of standing gravitational waves and the
associated power spectrum oscillations, versus traveling gravitational 
waves with no power spectrum oscillations, it was explicitely 
shown \cite{bg} that the later hypothesis does not produce oscillations 
in the $l$-space. 

Incidentally, it was argued \cite{bg} that in 
the case of density perturbations, the main contribution to the 
peaks in the temperature function $l(l+1)C_l$ can also be provided by 
oscillations in the metric power spectrum, rather than by the
temperature variations accompanying sound waves in the 
photon-electron-baryon plasma at the last scattering surface. In
the case of density perturbations, the metric power spectrum is
mostly associated with the gravitational field of the dark matter,
which dominates other matter components in terms of the gravitational
field. 

Oscillations in the metric power spectrum in the early
Universe are inevitable, and for the same reason as in the g.w.\
case, namely, because of the standing-wave pattern of the metric 
perturbations that is related to their quantum-mechanical origin. Therefore, 
the often-discussed ``acoustic" peaks in the $l$-space may well
turn out to be ``gravitational" peaks. It remains to be seen how
this circumstance can change inferences about cosmological parameters. 

We shall now turn to the CMB polarisation. (For some important papers
on CMB polarisation, see for example \cite{r, bp, pol, be, sz, kks, hu}.) 
It follows from the radiation transfer equations that the 
polarisation of CMB is mainly determined by the first time-derivative of the 
metric perturbations in the interval of time when the polarisation 
is mainly produced. Therefore,
it is the power spectrum of the function $h^{\prime}_{ij}(\eta, \bf{x})$
that is of a primary importance. Since the g.w.\ field itself, including
its normalisation, has been fully determined, the quantity of our 
interest is directly calculable. In Fig. 7 we show \cite{bgp} the power 
spectrum $(h^{\prime}/n)_{rms}^2(n)$, calculated at the
time of decoupling. 
\begin{figure}[!hbt] 
\caption{CMB temperature angular power spectrum} 
\centerline{\psfig{file=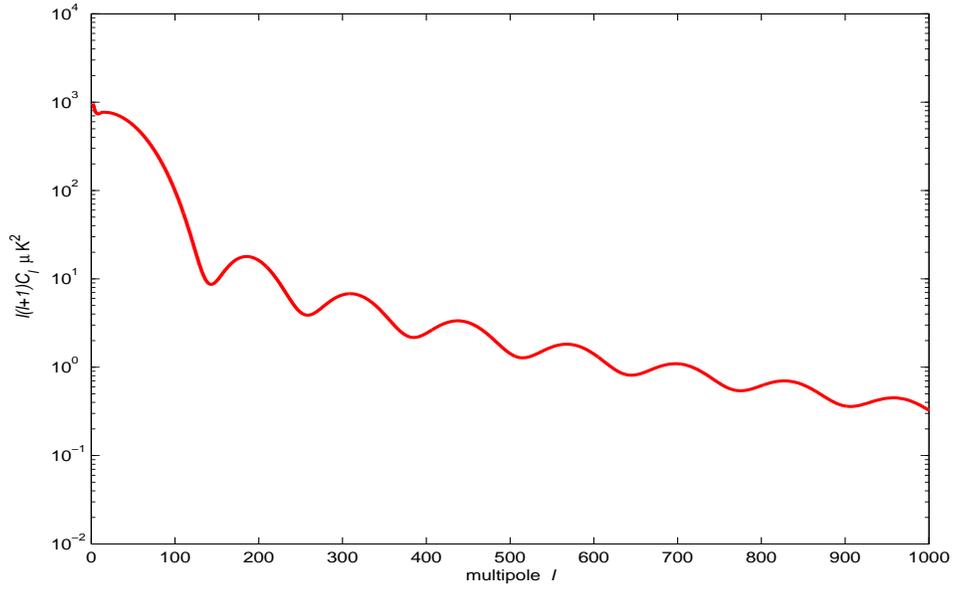,height=3.1in,width=5in}}
\end{figure}
\begin{figure}[!hbt] 
\caption{Power spectrum of gravitational waves at decoupling} 
\centerline{\psfig{file=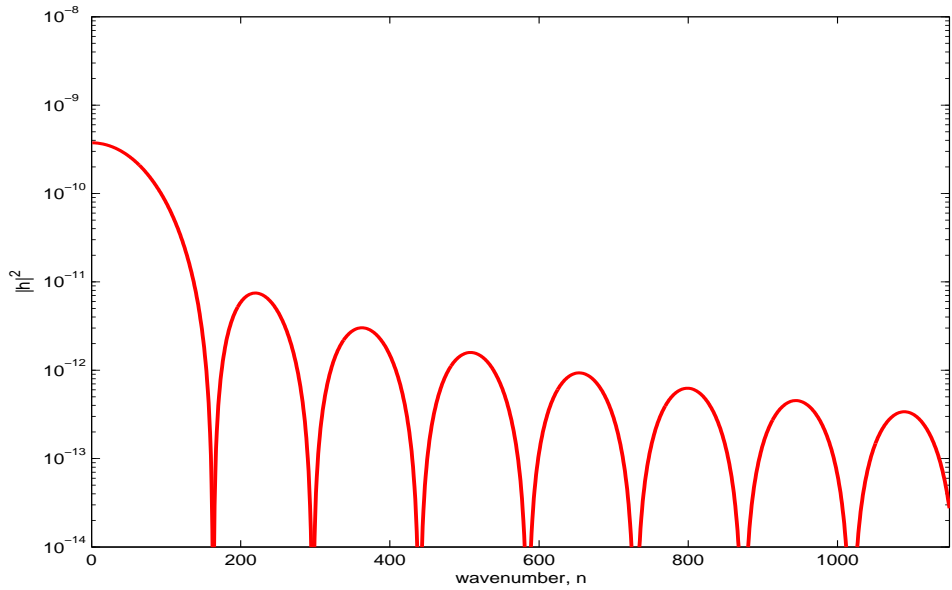,height=3.1in,width=5in}}
\end{figure}
\begin{figure}[!hbt] 
\caption{Angular power spectrum for CMB polarisation} 
\centerline{\psfig{file=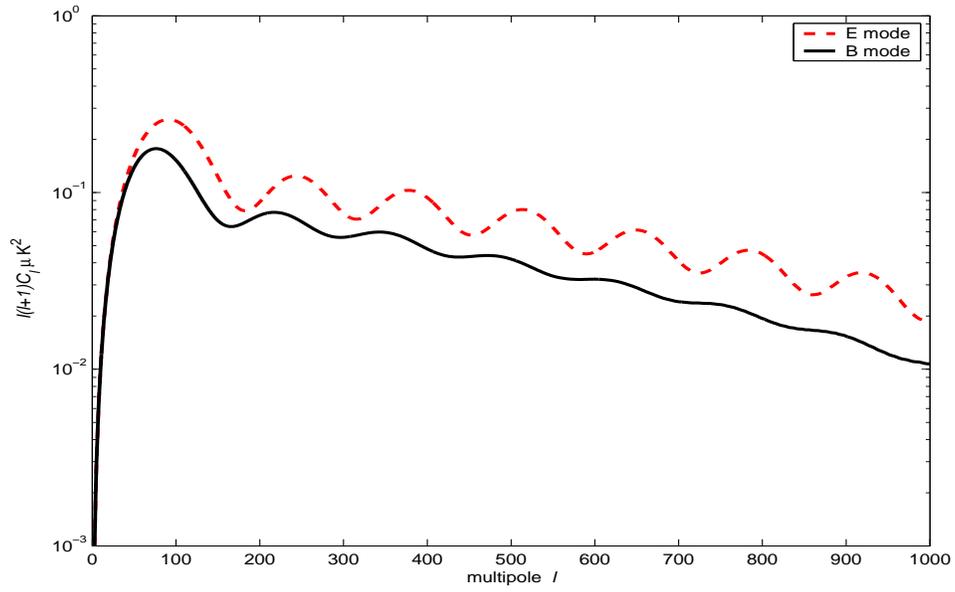,height=3.1in,width=5in}}
\end{figure}
\begin{figure}[!hbt] 
\caption{Power spectrum of first derivative of the g.w.\ metric} 
\centerline{\psfig{file=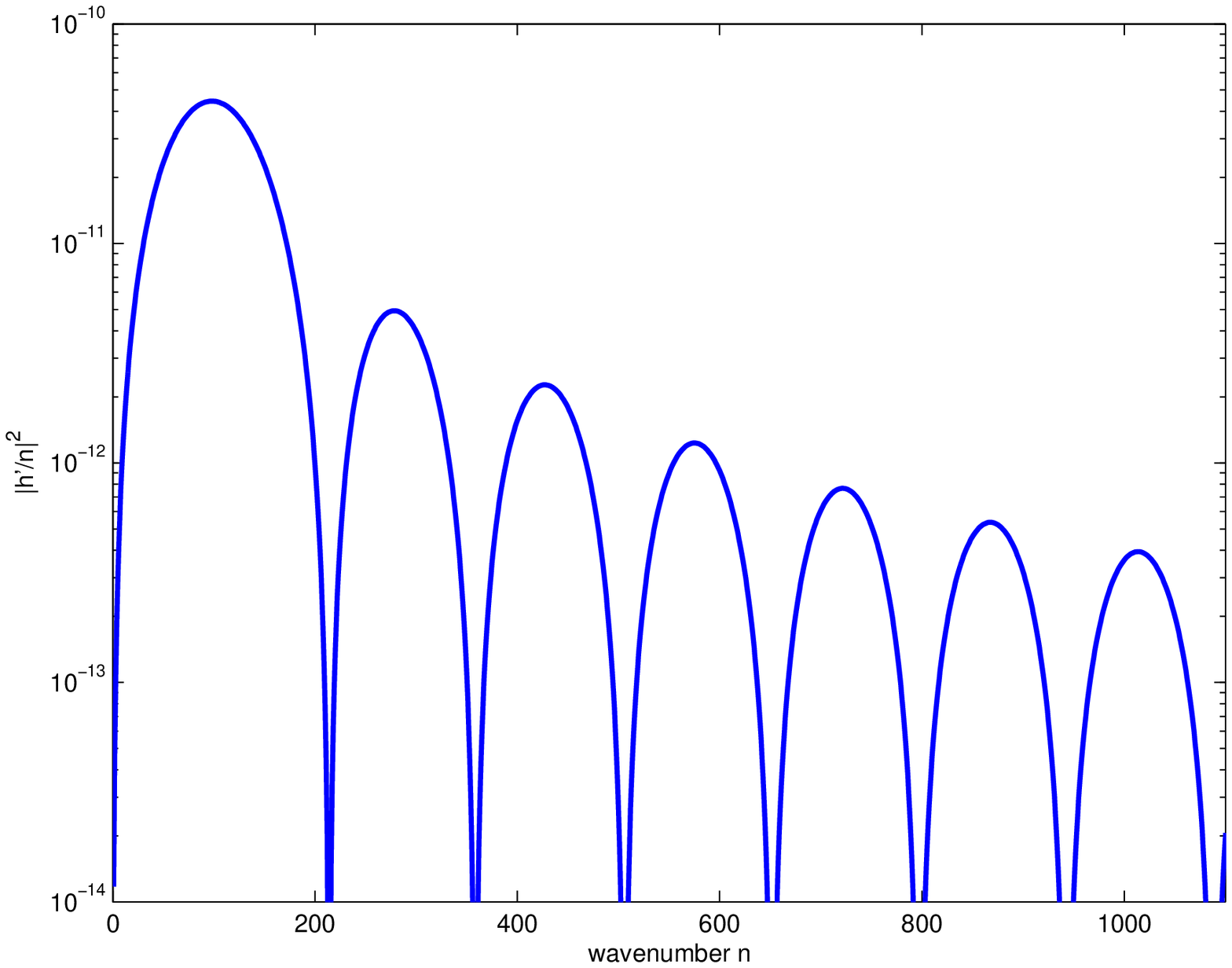,height=3.1in,width=5in}}
\end{figure}
The induced $E$ and $B$ components of polarisation
are shown in Fig. 6. This graph was derived under the usual assumptions
about the recombination history, which means, in particular, that the 
polarisation was primarily accumulated during a relatively short 
interval of time around $z_{dec}$. 

Similarly to the case of temperature 
anisotropies, the extrema in the graphs of Fig. 7 and Fig. 6 are 
in a good correspondence with each other. If there is not much power 
in the first time-derivative of the metric, you should not expect much 
power in the polarisation at the corresponding angular scale. On the 
other hand, the region of wavenumbers $n \approx 90$, where there is 
the first pronounced peak in Fig. 7, is fully responsible for the first 
pronounced peak in Fig. 6 at the corresponding angular 
scales $l\approx 90$.
\begin{figure}[!hbt] 
\caption{Expected numerical level of anisotropy and polarisation induced
by relic gravitational waves} 
\centerline{\psfig{file=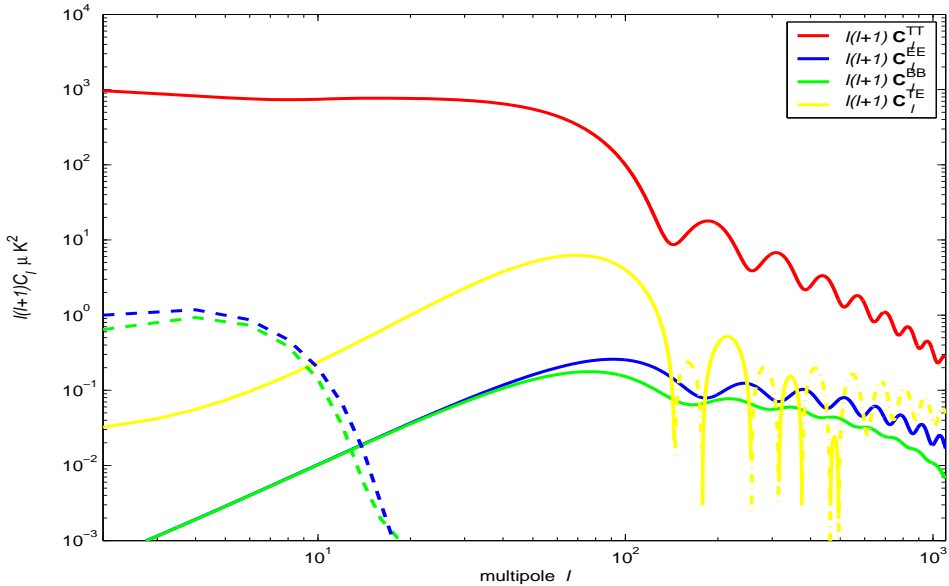,height=3.1in,width=5in}}
\end{figure}

In Fig. 8 we combine together some of the expected signals from 
relic gravitational waves. They are encoded in the CMB anisotropies 
and polarisation. 
This figure includes also a possible polarisation bump, discussed 
previously by other authors, at very small $l$'s. This feature arises 
because of the extended reionisation period in the relatively late 
universe, around $z_{rei} \approx 17$. In agreement 
with the explanations given above, the amplitude and position of this
bump in the $l$-space are determined by the amplitude and position
of the first maximum in the power spectrum $(h^{\prime}/n)^2$ of 
the function $h^{\prime}_{ij}(\eta, \bf{x})$, calculated at $z_{rei}$. 

The resulting graphs in Fig. 6 and Fig. 8 are qualitatively 
similar to the graphs derived by other authors before us. However, 
we take the responsibility of claiming that the numerical level of,
say, the $B$ component of polarisation shown in our graphs is what the 
observers should expect to see on the sky. Of course, this statement 
assumes that the observed large-scale anisotropies of CMB are 
caused by cosmological perturbations of quantum-mechanical origin, 
and not by something else. 

The true level of the $B$ signal can be 
somewhat higher or somewhat lower than the theoretical level shown in our 
figures. But the signal cannot be, say, several orders of magnitude 
lower than the one shown on our graphs. In contrast, the
inflationary literature claims that the amount of ``inflationary
gravitational waves" vanishes in the limit of the 
flat primordial spectrum $\beta = -2$ (${\rm n}=1$). Therefore, the
most likely level of the $B$ mode signal produced by ``inflationary 
gravitational waves" is close to zero. This would make the detection 
impossible in any foreseeable future. It is a pity 
that many of our experimentall colleagues, being guided by the wrong 
theory, are accepting their defeat even before 
having started to build instruments aimed at detecting 
relic gravitational waves via the $B$ component of polarisation. 

Their logic seems to be the following: `we would love to discover 
the fundamentally important relic gravitational waves, but we were 
told by inflationists many times that this is very unlikely to
happen, so we agreed to feel satisfied even if we succeed
only in putting some limits on, say, polarisation properties of
dust in the surrounding cosmos'. The author of this contribution
fears that in a complex experiment like the $B$-mode detection, 
this kind of logic can only lead to overlooking the 
important signal that the experiment originally targeted.

Concluding this section I would like to say as a witness 
that Zeldovich suggested using the CMB polarisation 
as a g.w.\ discriminator, as early as in the very beginning 
of the 80's. This was clearly stated in private conversations, 
but I am not aware of any written records. 

\section{The false ``standard inflationary result". How to
correctly quantise a cosmological harmonic oscillator} 

Why bother about relic gravitational waves if inflationists claim 
that the amount of relic gravitational waves (inflationists and
followers call them ``inflationary gravitational waves") should be 
zero or almost zero?  This claim is a direct consequence of the 
so-called ``standard inflationary result", which is the main 
contribution of inflationary theorists to the subject of practical,
rather than imaginary, cosmology. 

In the inflationary scenario, the `initial' era of the universe 
expansion is driven by a scalar field $\varphi$ with the scalar 
field potential $V(\varphi)$. It is in this era that the initial 
quantum vacuum conditions for cosmological pertubations are being 
formulated. The inflationary solutions for the scale factor $a(\eta)$ 
are close to the deSitter evolution characterised 
by $\beta =-2$ in Eq.(\ref{a}). The effective equation 
of state for the scalar field is always $\epsilon +p \geq 0$,
so that for the power-law intervals of expansion driven by the scalar 
field, the parameter $\beta$
can only be $\beta \leq -2$, see Eq.(\ref{w}). Therefore, one 
expects the primordial spectrum of the generated metric perturbations
to be almost flat, i.e. the primordial spectral index ${\rm n}$ 
should be close to ${\rm n} =1$, with ${\rm n} \leq 1$.

The beginning of the amplifying 
superadiabatic regime for the given mode of perturbations is often
called the `first Hubble radius crossing', while the end of 
this regime for the given mode is often called the `second Hubble 
radius crossing'. The ``standard inflationary result" is formulated for
cosmological perturbations called density perturbations 
(scalar, $S$, perturbations) as opposed to the
gravitational waves (tensor, $T$, perturbations) considered in Sec. 1. 

The ``standard inflationary result" states that the final (second
crossing, $\bar f$) amplitudes of quantities describing density 
perturbations are related
to the initial (first crossing, $i$) values of $\varphi$ and other 
quantities, according to the evaluation:
\begin{equation}
\label{sir}
\left(\frac{\delta \rho}{\rho}\right)_{\bar f} \sim \left(h_S\right)_{\bar f} 
\sim \left(\zeta \right)_{\bar f} \approx
\left(\zeta \right)_{i} \sim \left(\frac{H^2}{\dot \varphi} \right)_{i} 
\sim \left( \frac{V^{3/2}}{V_{,\varphi}}\right)_{i} 
\sim \frac{H_{i}}{\sqrt{1-{\rm n}}}.
\end{equation}

The numerator of the last term on the r.h.s. of Eq.(\ref{sir}) is the
value of the Hubble parameter taken at the moment of time when the given
mode enters the superadiabatic regime. This is the same quantity $H_i$
which defines the g.w.\ (`tensor') metric amplitude, as described 
in Sec. 1. Since we are supposed to start with the initial vacuum
quantum state for all cosmological perturbations, one would expect 
that the results
for density perturbations should be similar to the results for
gravitational waves. One would expect that the amplitude $h_S$ 
of the generated `scalar' metric perturbations should be  
finite and small, and of the same order of magnitude as the 
amplitude $h_T$ of `tensor' metric perturbations. 

However, according
to the ``standard inflationary result", this is very far from being
the case. The denominator of the last term of Eq.(\ref{sir}) contains
a new factor: $\sqrt{1 -{\rm n}}$. This factor goes to zero in the 
limit of the most interesting and observationally preferred 
possibility of the flat (Harrison-Zeldovich-Peebles) primordial 
spectrum ${\rm n} =1$. Correspondingly, the amplitudes of the 
generated density perturbations go to infinity, according
to the prediction of inflationary theorists, in the limit of 
the flat spectrum. (By now, the ``standard inflationary result" 
(\ref{sir}) has been cited, used, praised, reformulated, 
popularised, etc. in hundreds of inflationary publications, 
so it has become `accepted by way of repetition'.) 

As will be demonstrated below, the divergence in Eq.(\ref{sir}) is 
not a violation, suddenly descending upon us from the `blue sky', 
of the adopted approximation of small linear perturbations. This 
is a manifestation of the incorrect theory. Even if the spectral
index ${\rm n}$ is not very close to 1, and you combine ${\rm n}$
with a reasonable $H_{i}$ in order to obtain, for example, 
a small number $10^{-5}$ for the r.h.s.\ of 
Eq.(\ref{sir}), this will not make your theory correct. This will 
be just an acceptable number accidentally following from the wrong 
formula. You will have to pay a heavy price in some other places.

An attempt to derive physical conclusions from this formula can 
only lead to mistakes. The current literature is full of
incorrect far-reaching physical conclusions derived from this 
wrong theory. This is a kind of situation which L. D. Landau used 
to describe sarcastically in the following words: ``If you 
assume that the derivative of the function $\sin x$ is $\ln x$, 
rather than $\cos x$, you can make many wonderful discoveries....".

In inflationary literature, the `zero in the denominator' factor
$\sqrt{1-{\rm n}}$ appears in many different dresses. It is often 
written in equivalent forms, such as $\left(\dot{\varphi}/ H\right)_{i}$, 
$\left(V_{, \varphi}/V \right)_{i}$, 
$\left(H_{, \varphi}/H \right)_{i}$, $\sqrt{1 +w_{i}}$, etc.     
Inflationists are routinely hiding their absurd prediction of infinitely 
large amplitudes of density perturbations that should take place 
in the limit of the flat spectrum ${\rm n} \rightarrow 1$. They divide 
the g.w.\ amplitude $h_T$ over the predicted divergent 
amplitude $h_S$. This division produces the so-called
`tensor-to-scalar ratio', or `consistency relation': 
$ h_T/h_S \approx \sqrt{1-{\rm n}}$. The quantity $H_i$,
common for the $T$ and $S$ perturbations, cancels
out in the composed ratio, and the `zero in the denominator'
factor is transferred to the numerator of the final expression.
It is then declared that the metric amplitude $h_S$ of density
perturbations is determined by the observed CMB anisotropies, and, 
therefore, the inflationary `consistency relation' demands that 
the g.w.\ amplitude $h_T$ must vanish in the 
limit of ${\rm n} \rightarrow 1$. 

In other words, instead of being
horrified by the fact that their theory predicts arbitrarily large 
amplitudes of density perurbations (and, hence, the theory is in complete
disagreement with observations, because the analysis of the data
shows no catastrophic increase in the amplitude when the tested
spectral index approaches ${\rm n}=1$), supporters of the inflationary
approach to science systematically claim that their theory is in 
`spectacular agreement' with observations,
and it is gravitational waves that should vanish. 

If this were true, there would not be much sense in attempting 
to detect primordial gravitational waves, as the 
observations persistently point toward ${\rm n} \approx 1$,
including ${\rm n} =1$. It is quite common to hear these days
enthusiastic promises of inflationary believers 
to detect ``inflationary gravitational waves" in the 
``not-so-distant future" via the measurement of $B$-mode polarisation 
of CMB. But from other papers of the same authors it follows 
that there is no reason even to try. If you trust and cite
inflationary formulas, the expected amount of ``inflationary 
gravitational waves" should be very small or zero. You can only 
hope to be extremely lucky if you suggest to detect them, even in 
the quite distant future, for example with the proposed mission
called Big Bang Observer. And nobody should be surprised if you have found 
nothing, because ${\rm n} =1$ is in the heart of all claims, 
theoretical and observational. Moreover, most loyal inflationists 
would say that this was exactly what they had always been 
predicting.

To demonstrate the incorrectness of inflationary conclusions, we 
shall now concentrate on the `zero in the denominator' factor. 
We will have to recall the quantisation procedure for 
gravitational waves and density perturbations. It is necessary
to remind the reader that some inflationists and their supporters 
insisted for many years on the claim that the dramatic difference 
in the final numerical values of $h_T$ and $h_S$ arises not because 
of the initial conditions, but because of the subsequent evolution. 

Specifically, they claimed that the classical 
long-wavelength `scalar' metric perturbations are capable of 
experiencing, in contrast to gravitational waves, 
a ``big amplification during reheating". (For a critical 
discussion, see \cite{g5}.) But it now 
looks as if the fallacy of this proposition has become clear 
even to its most ardent proponents. Therefore, we shall now 
focus on the issue of quantum mechanics and initial conditions.

The perturbed gravitational field for all three sorts of 
cosmological perturbations (scalar, vector, tensor) is described
by Eq.(\ref{metric}). For simplicity, we are considering 
spatially flat cosmologies, whose spatial curvature radius is
infinite. However, if the spatial curvature radius is finite but, 
say, only a factor of 10 longer than $l_H$, very little will 
change in our analysis. 

The metric perturbations $h_{ij}(\eta, {\bf x})$ can be 
expanded over spatial Fourier harmonics labeled by the
wavevector ${\bf n}$:
\begin{eqnarray}
\label{hij}
& &h_{ij}(\eta,{\bf x}) =  \\  \nonumber 
& &\frac{\cal C}{(2\pi )^{3/2}} \int_{-\infty}^{\infty} d^3{\bf n}
\sum_{s=1, 2}~{\stackrel{s}{p}}_{ij} ({\bf n}) \frac{1}{\sqrt{2n}}
\left[ {\stackrel{s}{h}}_n (\eta ) e^{i{\bf n}\cdot {\bf x}}~
                 {\stackrel{s}{c}}_{\bf n}
                +{\stackrel{s}{h}}_n^{\ast}(\eta) e^{-i{\bf n}\cdot {\bf x}}~
                 {\stackrel{s}{c}}_{\bf n}^{\dag}  \right]. 
\end{eqnarray}
The three sorts of cosmological perturbations are different in that
they have three different sorts of polarisation tensors 
${\stackrel{s}{p}}_{ij} ({\bf n})$, and each of them has two
different polarisation states $s= 1, 2$. The `scalar' and `vector'
metric perturbations are always accompanied by perturbations in density
and/or velocity of matter. The normalisation constant
${\cal C}$ is determined by quantum mechanics, and the derivation
of its value is one of the aims of our discussion.

Let us recall the procedure of quantisation of gravitational waves. 
Let us consider 
an individual g.w. mode ${\bf n}$. The time-dependent mode functions 
${\stackrel{s}{h}}_n (\eta )$ can be written as
\begin{equation}
\label{mf}
{\stackrel{s}{h}}_n (\eta )= \frac{1}{a(\eta)}{\stackrel{s}{\mu}}_n (\eta ).
\end{equation}
For each $s$ and ${\bf n}$, the g.w.\ mode functions $\mu(\eta)$ satisfy the
familiar equation~(\ref{mu}).

The action for each mode has the form    
\begin{equation}
\label{actg}
S = \int L~ d \eta,
\end{equation}
where the g.w. Lagrangian $L$ is given by the expression \cite{g6}
\begin{equation}
\label{Lg}
L_{gw} = \frac{1}{2c \kappa}n^{-3}a^2\left[{\left(
\frac{\mu}{a}\right)^{\prime}}^2 -
n^2 \left(\frac{\mu}{a}\right)^2\right], 
\end{equation}
and
\[
\kappa = \frac{8\pi G}{c^4}.
\]

The Euler-Lagrange equation 
\[
\frac{\partial L}{\partial h} -\frac{d}{d \eta} 
\frac{\partial L}{\partial h^{\prime}} = 0
\]
for the dimensionless g.w.\ variable $h= \mu/a$ brings us to
the equation of motion
\begin{equation}
\label{heq}
h^{\prime \prime} +2\frac{a^{\prime}}{a} h^{\prime}+n^2 h=0,
\end{equation}
which is equivalent to Eq.(\ref{mu}). 

In order to move from 3-dimensional Fourier components to the
usual description in terms of an individual oscillator with
frequency $n$, we will be working with the quantity  
$\bar{h}$ introduced according to the definition
\begin{equation}
\label{hbar}
\bar{h} = \frac{a_0}{n \sqrt{c\kappa}} h = \frac{\sqrt{\hbar} a_0}
{\sqrt{8\pi} l_{Pl} n} h = 
\sqrt{\frac{\hbar}{32 \pi^3}} \frac{\lambda_0}{l_{Pl}} h,
\end{equation}
where $a_0$ is a constant. This constant $a_0$ is the value of the 
scale factor $a(\eta)$ at some instant of time $\eta = \eta_0$ where the 
initial conditions are being formulated, 
and $\lambda_0 = 2\pi a_0/n$. 

In terms of $\bar{h}$, 
the Lagrangian (\ref{Lg}) takes the form 
\begin{equation}
\label{Lg2}
L_{gw} = \frac{1}{2n} \left(\frac{a}{a_0}\right)^2 
\left[\left( {\bar{h}}'\right)^2 -n^2 {\bar{h}}^2\right]. 
\end{equation}

The quantity $\bar{h}=q$ is the `position' variable, while the canonically 
conjugate `momentum' variable $p$ is
\begin{equation}
\label{mom}
p = \frac{\partial L}{\partial {\bar{h}}^{\prime}} =\frac{1}{n}
\left(\frac{a}{a_0}\right)^2 {\bar{h}}^{\prime}.
\end{equation} 

In the distant past, at times near $\eta_0$, and before $\eta_i$ when
a given mode entered superadiabatic regime, the g.w.\ amplitude behaved 
according to the law  
\[
h(\eta) \propto \frac{1}{a(\eta))} e^{-in(\eta - \eta_0)}. 
\]
The time-derivative
of $a(\eta)$ can be neglected, i.e. $a^{\prime}/a \ll n$. Then, we 
promote $q$ and $p$ to the status of quantum-mechanical operators, 
denote them by bold-face letters, and write down their asymptotic 
expressions:
\begin{equation}
\label{qp1}
{\bf q}= \sqrt{\frac{\hbar}{2}} \frac{a_0}{a}\left[{\bf c}e^{-in(\eta-\eta_0)} 
+ {\bf c}^{\dagger}e^{in(\eta-\eta_0)} \right],
\end{equation}
\begin{equation}
\label{qp2}
{\bf p}= i\sqrt{\frac{\hbar}{2}} \frac{a}{a_0}\left[-{\bf c}
e^{-in(\eta-\eta_0)} + {\bf c}^{\dagger}e^{in(\eta-\eta_0)} \right].
\end{equation}

The commutation relationships for the ${\bf q}, {\bf p}$ operators, and
for the annihilation and creation operators ${\bf c}, {\bf c}^{\dagger}$, 
are given by
\[
\left[{\bf q},{\bf p} \right] =i\hbar, ~~~~~~~~~~~\left[ {\bf c}, 
{\bf c}^{\dagger}\right]=1.
\]
The initial vacuum state $|0\rangle$ is defined by the condition
\[
{\bf c}|0\rangle =0.
\]
This is indeed a genuine vacuum state of a simple harmonic
oscillator, which gives at  $\eta = \eta_0$ the following relationships
\[
\langle 0|{\bf q}^2|0 \rangle = \langle 0|{\bf p}^2|0 \rangle =\frac{\hbar}{2},
~~~~~~~~\Delta {\bf q} \Delta {\bf p} =\frac{\hbar}{2}.
\]

The root-mean-square value of ${\bf q}$ in the vacuum state 
is $q_{rms} = \sqrt{\hbar/2}$.
Combining this number with the definition (\ref{hbar}) we derive
\begin{equation}
\label{hrmsgw}
h_{rms} = {\left(\langle 0|{\bf h}^2|0 \rangle\right)}^{1/2} =
\frac{\sqrt{2}(2 \pi)^{3/2} l_{Pl}}{\lambda_0}.
\end{equation}
Extrapolating the initial time $\eta_0$ up to the boundary between the
adiabatic and superadiabatic regimes at $\eta =\eta_i$, we derive the 
estimate $h_{rms} \sim l_{Pl}/ \lambda_i$. It is this evaluation
that was used in \cite{g1} and in Sec. 1. More accurate calculations 
along these lines produce ${\cal C} = \sqrt{16 \pi} l_{Pl}$ in
expression (\ref{hij}) for gravitational waves. 

A consistent formal derivation of the total Hamiltonian, including
the terms describing interaction of the oscillator with external field,
is presented in Ref.\cite{grcqg} by equations
(19)-(24) there. Technically,
the derivation is based on the canonical pair $q=\mu$, 
$p = \partial L/\partial \mu^{\prime}$. The Hamiltonian associated with
the Lagrangian (\ref{Lg2}) has the form 
\begin{equation}
\label{hamgw}
{\bf H}(\eta) = n {\bf c}^{\dagger}{\bf c} + \sigma {{\bf c}^{\dagger}}^2 +
\sigma^{*} {\bf c}^2,
\end{equation}
where the coupling to the external field is given by the function
$\sigma(\eta) = (i/2)(a^{\prime}/a)$. In the same Ref.\cite{grcqg} one can
also find the Heisenberg equations of motion for the Heisenberg operators
${\bf c}(\eta)$, ${\bf c}^{\dagger}(\eta)$, and their connection to 
classical equation (\ref{mu}). The asymptotic expressions for the
Heisenberg operators,
\[
{\bf c}(\eta) = {\bf c} e^{-in(\eta- \eta_0)}, ~~~~~
{\bf c}^{\dagger}(\eta) = {\bf c}^{\dagger} e^{in(\eta-\eta_0)},
\] 
enter into formulas (\ref{qp1}), (\ref{qp2}). Clearly, the vacuum state
$|0\rangle$, defined as ${\bf c}(\eta)|0\rangle =0$, minimises the
oscillator's energy (\ref{hamgw}).

A rigorous quantum-mechanical Schrodinger
evolution of the initial vacuum state of cosmological perturbations 
transforms this state into a strongly squeezed (multi-particle) vacuum
state \cite{g6}, but we focus here only on the initial quantum state, which 
defines the quantum-mechanical normalisation of our classical mode functions.

\vspace{0.5cm}

We shall now switch to density perturbations.

\vspace{0.5cm}

For each mode $\bf n$ of density perturbations (S-perturbations), 
the mode's metric 
components $h_{ij}$ entering Eq.(\ref{metric}) can be written as 
\[
h_{ij}= h(\eta) Q \delta_{ij} + h_l(\eta) n^{-2} Q_{,ij},
\]
where the spatial eigen-functions $Q$ are $Q = e^{\pm i{\bf n}\cdot{\bf x}}$. 
Therefore, the metric components associated with density 
perturbations are characterised by two polarisation 
amplitudes: $h(\eta)$ and $h_l(\eta)$. If the initial era is driven 
by an arbitrary scalar field $\varphi$,
there appears a third unknown function - the amplitude 
$\varphi_1(\eta)$ of the scalar field perturbation:
\[
\varphi = \varphi_0(\eta) + \varphi_1(\eta) Q. 
\]

One often considers
the so-called minimally coupled to gravity scalar field $\varphi$, 
with the energy-momentum tensor
\[
T_{\mu \nu} = \varphi_{,\mu} \varphi_{,\nu} - g_{\mu \nu} \left[
\frac{1}{2}g^{\alpha \beta} \varphi_{,\alpha} \varphi_{,\beta} + 
V(\varphi) \right].
\]
The coupling of scalar fields to gravity is still a matter of
ambiguity, and the very possibility of quantum-mechanical 
generation of density perturbations relies on an extra hypothesis, 
but we suppose that we were lucky and the coupling was such as we need.
The three unknown functions $h(\eta)$, $h_l(\eta)$, $\varphi_1(\eta)$ 
should be found 
from the perturbed Einstein equations augmented by the appropriate 
initial conditions dictated by quantum mechanics. 

It is important to note that inflationary theorists are still
struggling with the basic equations for density perturbations.
In inflationary papers, you will often see equations containing 
complicated combinations of metric perturbations mixed up 
with the unperturbed and/or perturbed functions of the scalar
field $\varphi$ and $V(\varphi)$. 

Inflationists are still engaged 
in endless discussions on the shape of the scalar field potential
$V(\varphi)$ and what it could mean for countless inflationary 
models. However, this state of affairs is simply a reflection of 
the fact that the equations have not been properly transformed and 
simplified. Since the underlying physics is the interaction of a 
cosmological harmonic oscillator with the gravitational pump field, 
mathematically the equations should reveal this themselves. 
And indeed they do.
 
It was shown in paper \cite{g7} that, for any potential $V(\varphi)$, 
there exists 
only one second-order differential equation to be solved:
\begin{equation}
\label{dpeq}
\mu^{\prime\prime} + \mu \left[n^2 - \frac{(a \sqrt{\gamma})^{\prime\prime}}
{a \sqrt{\gamma}}\right] = 0,
\end{equation}
where the function $\mu(\eta)$ represents the single dynamical degree of 
freedom, describing S-perturbations. The effective potential barrier
$(a \sqrt{\gamma})^{\prime\prime}/(a \sqrt{\gamma})$
depends only on $a(\eta)$ and its derivatives, in full analogy
with the g.w.\ oscillator, Eq.(\ref{mu}). The time-dependent 
function $\gamma$ ($\gamma(\eta)$ or $\gamma(t)$) is defined by
\[
\gamma = 1 + \left(\frac{a}{a^{\prime}} \right) ^{\prime} =
- \frac{c}{a} \frac{H^{\prime}}{H^2}= - \frac{\dot H}{H^2}.
\]

As soon as the appropriate solution for $\mu(\eta)$ is found, all 
three functions describing density perturbations are easily calculable:
\[
h(\eta) = \frac {1}{c} H(\eta)\left[ \int_{\eta_0}^{\eta} \mu \sqrt{\gamma}
d\eta +C_i\right],
\]
\[
{h_l}^{\prime}(\eta)= \frac{a}{a'} \left[h''- \frac{H''}{H'} h' +n^2 h \right],
\]
\[
\varphi_1(\eta) = \frac{\sqrt \gamma}{\sqrt{2 \kappa}} \left[\frac{\mu}
{a \sqrt \gamma} - h \right].
\]
The constant $C_i$ reflects the remaining coordinate freedom within the
class of synchronous coordinate systems. (Another constant comes out
from the integration of the above-given equation for ${h_l}^{\prime}$.) 
The funcion $\mu$ does 
not depend on this remaining coordinate freedom, and the constant $C_i$ 
cancels out in the expression defining $\mu(\eta)$ in terms of $h(\eta)$:
\[
\frac{\mu}{a \sqrt \gamma} =  h- \frac{H}{H'} h'.
\]
The function $\mu/a\sqrt \gamma$ is that part of the scalar metric
amplitude $h(\eta)$ which does not depend on the remaining 
coordinate freedom (`gauge-invariant' metric perturbation).

In the short-wavelength regime, the function $\mu$ describing density 
perturbations behaves as $\mu \propto e^{-in\eta}$. This is the same 
behaviour as in the case of the function $\mu$ describing gravitational
waves. This similarity between the respective functions $\mu$~
($\mu_T$ and $\mu_S$) is valid 
only in the sense of their asymptotic $\eta$-time dependence, 
but not in the sense of their overall numerical normalisation (see below). 

In the long-wavelength regime, the dominant solution to Eq.(\ref{dpeq}) is 
$\mu \propto a \sqrt \gamma$. The quantity which remains constant 
in this regime is $\mu /a \sqrt\gamma$. It is this physically relevant 
variable that takes over from the analogous variable $h=\mu/a$ in the 
g.w.\ problem. We introduce the notation
\begin{equation}
\label{zeta}
\frac{\mu}{a \sqrt \gamma} = \zeta,
\end{equation}
where $\mu$ satisfies Eq.(\ref{dpeq}).

To make contact with earlier work, it should be mentioned that the
previously introduced quantity
\[
\zeta_{BST} = \frac{2}{3} \frac{(a/a') {\Phi}' +\Phi}{1+w} +\Phi,
\]
where $\Phi$ is Bardeen's potential and $BST$ stands for Bardeen,
Steinhardt, Turner \cite{bst}, can be reduced to our variable $\zeta$ 
(\ref{zeta}) up to the numerical coefficient $-(1/2)$. Our 
quantity $\mu$ for density perturbations can also be related to the 
variable $u_{CLMS}$, where $CLMS$ stands for Chibisov, Lukash, 
Mukhanov, Sasaki \cite{luk, chm, sas}.

In preparation for quantisation, we should first identify the
inflationary `zero in the denominator' factor. The unperturbed
Einstein equations for the coupled system of gravitational
and scalar fields require \cite{g7}
\[
\kappa \left({\varphi_0}^{\prime}\right)^2 =2 \left(\frac{a'}{a}\right)^2
\gamma.
\]
Therefore, the `zero in the denominator' factor
\[
\left( \frac{\dot{\varphi}_0}{H}\right)_i =\sqrt{\frac{2}{\kappa}}
\left(\sqrt{\gamma} \right)_i
\]
is expressed in the form of very small values of the dimensionless 
function $\sqrt \gamma$. 

Within the approximation
of power-law scale factors (\ref{a}), the function $\gamma$ reduces to 
a set of constants,
\[
\gamma = \frac{2+\beta}{1+\beta}, ~~~~1+w = \frac{2}{3} \gamma.
\]
The constant $\gamma$ degenerates to zero in the limit of the evolution 
law with $\beta = -2$; that is, in the limit of the gravitational
pump field which is responsible for the generation of primordial 
cosmological perturbations with flat spectrum ${\rm n} =1$. So, we are 
especially interested in the very small values of $\sqrt \gamma$. 

It was shown \cite{g7} that the dynamical problem for the 
scalar-field-driven S-perturbations can be obtained 
from the dynamical problem for gravitational waves by simple
substitutions: $a(\eta) \rightarrow a(\eta) \sqrt{\gamma(\eta)}$,~
$\mu_T(\eta) \rightarrow \mu_S(\eta)$. 
(This is not a conjecture, but this is a rule whose validity was established
after a thorough analysis of these two problems separately.) 
Each of these substitutions is valid up to an arbitrary constant factor.
Using these substitutions, one obtains the
S-equation (\ref{dpeq}) from the T-equation (\ref{mu}),
and one obtains the physically relevant variable 
$\zeta=\mu_S/a\sqrt{\gamma}$ for S-perturbations from 
the g.w.\ variable $h= \mu_T/a$. 

Moving from the 3-dimensional Fourier components of the field $\zeta$
to an individual oscillator with frequency $n$, we introduce the
quantity $\bar{\zeta}$ according to the same rule (\ref{hbar}) that
was used when we introduced $\bar{h}$. Namely, we introduce
\begin{equation}
\label{zbar}
\bar{\zeta} = \frac{a_0}{n \sqrt{c\kappa}} \zeta = \frac{\sqrt{\hbar} a_0}
{\sqrt{8\pi} l_{Pl} n} \zeta =
\sqrt{\frac{\hbar}{32 \pi^3}} \frac{\lambda_0}{l_{Pl}} \zeta.
\end{equation}
The application of the substitutions 
$a \rightarrow \tilde{a} = a \sqrt\gamma$,
$\bar{h} \rightarrow \bar{\zeta}$ to the g.w.\ Lagrangian (\ref{Lg2}) gives 
rise to the Lagrangian $L_{dp}$ for the single dynamical degree of freedom 
describing S-perturbations:
\begin{equation}
\label{Ld2}
L_{dp} = \frac{1}{2 n}\left(\frac{a \sqrt \gamma}
{a_0 \sqrt{\gamma_0}} \right)^2\left[\left({\bar{\zeta}}^{\prime}\right)^2 -
n^2 {\bar{\zeta}}^2\right]. 
\end{equation}

Obviously, the Euler-Lagrange equation
\begin{equation}
\label{zheq}
\zeta^{\prime \prime} +2\frac{(a \sqrt \gamma)^{\prime}}{a \sqrt \gamma}, 
\zeta ^{\prime}+n^2 \zeta=0
\end{equation}
derivable from the lagrangian (\ref{Ld2}) in
terms of the independent variable $\zeta$,
is equivalent to Eq.(\ref{dpeq}) which is the Euler-Lagrange equation
derivable from the Lagrangian (\ref{Ld2}) in terms of the 
independent variable $\mu_S$. The Lagrangian (\ref{Ld2}) should be 
used for quantisation. The Lagrangian
itself, as well as the action and the Hamiltonian, does not degenerate in
the limit $\gamma \rightarrow 0$, i.e., in the limit of the most
interesting background gravitational field in the form of
the de-Sitter metric, $\gamma =0$.

We shall start with the analysis of the paper \cite{mal} 
which, together with Ref.\cite{mal2}, is sometimes referred to as 
the most recent work that 
contains a rigorous mathematical derivation of the ``standard   
inflationary result". The author of these papers uses slightly 
different notations, such as $a^2 = e^{2\rho}$ and $\varphi = \phi$. 
In his notation, the 
quantity $\dot{\varphi}_0/H$ is $\dot{\phi} /\dot{\rho}$,
so that the `zero in the denominator' factor appears as
$\dot{\phi}_{*} /\dot{\rho}_{*}$, where the asterisk means ``the
time of horizon crossing". 

As a ``useful example to keep 
in mind" for quantisation of density perturbations, the author 
suggests the artificial model of a test massless scalar field $f$ 
in deSitter space. But the Lagrangian, classical solutions,
and quantisation procedure for $f$ are identical to the g.w.\ case 
that we recalled above, so that his variable $f$ is our $h$ for 
gravitational waves. His Lagrangian (2.12) for density 
perturbations coincides in structure with our Lagrangian (\ref{Ld2}), 
and we discuss one and the same observable quantity $\zeta$. 

It is worthwhile to quote explicitely the attempted rigorous 
proof \cite{mal}: ``Since the action (2.12) also contains a factor 
$\dot{\phi} /\dot{\rho}$ we also have to set its value to the 
value at horizon crossing, this factor only appears in normalizing
the classical solution. In other words, near horizon crossing we
set 
\[f = \frac{\dot{\phi}}{\dot{\rho}} \zeta, 
\]
where $f$ is a canonically
normalized field in de-Sitter space. This produces the well known
result...". And the author immediately writes down the square of 
the ``standard inflationary result", with the square of the factor 
$\dot{\phi}_{*} /\dot{\rho}_{*}$ in the denominator of the final
expression. 

Let us try to traverse in practice the path to the
``well known result". (To be fair to the author, the derivation of 
the ``standard inflationary result" does not appear to be the main 
purpose of his paper \cite{mal}, so my criticism does not imply anything
about other statements of that paper.) The factor  
$\dot{\phi} /\dot{\rho}$ in (2.12) of the cited paper is our
factor $\sqrt \gamma$ in Eq.(\ref{Ld2}). It is recommended \cite{mal} 
to combine the results for the g.w.\ variable $h$ with the prescription
$\zeta = \frac{1}{\sqrt \gamma}h$. So, instead of Eq.(\ref{qp1}),
we would have to write
\begin{equation}
\label{qp1d}
{\bf q}=\bar{\zeta}= \sqrt{\frac{\hbar}{2}} \frac{\tilde{a}_0}
{\tilde{a}} \frac{1}{\sqrt\gamma}\left[{\bf b}e^{-in(\eta-\eta_0)} 
+ {\bf b}^{\dagger}e^{in(\eta-\eta_0)} \right].
\end{equation}

The canonically conjugate momentum seems to be
\begin{equation}
\label{momd}
p = \frac{\partial L}{\partial {\bar{\zeta}}^{\prime}} =\frac{1}{n}
\left(\frac{\tilde{a}}{\tilde{a}_0}\right)^2 \gamma {\bar{\zeta}}^{\prime}.
\end{equation} 
The time derivative of $\gamma$ should be neglected, as $\gamma$ 
is either a constant or a slowly changing function at times near 
$\eta_0$. Therefore, we would have to write, instead of Eq.(\ref{qp2}),
the following relationship: 
\begin{equation}
\label{qp2d}
{\bf p}= i\sqrt{\frac{\hbar}{2}} \frac{\tilde{a}}{\tilde{a}_0} \sqrt\gamma
\left[-{\bf b}e^{-in(\eta-\eta_0)} + 
{\bf b}^{\dagger}e^{in(\eta-\eta_0)} \right].
\end{equation}
The commutation relations are given by
\[
\left[{\bf q},{\bf p} \right] =i\hbar, ~~~~~~~~~~~\left[ {\bf b}, 
{\bf b}^{\dagger}\right]=1.
\]

One is encouraged and tempted to think that the quantum 
state $|0_s\rangle$, annihilated by ${\bf b}$, namely
\[
{\bf b}|0_s \rangle =0,
\]
is the vacuum state of the field $\zeta$, i.e., the ground state of
the Hamiltonian associated with the Lagrangian (\ref{Ld2}). 
The calculation of the 
mean square value of $\bar{\zeta}$ at $\eta=\eta_0$ produces the result    
\[
\langle 0_s|{\bf q}^2|0_s \rangle = \frac{\hbar}{2} \frac{1}{\gamma_0},
\]
in which the `zero in the denominator' factor $\sqrt \gamma$ is manifestly
present and squared, as the ``well known result" prescribes. 

In the limit
of very small $\sqrt \gamma$ one obtains the divergence of initial
amplitudes, which is in the heart of all inflationary predictions. (In 
the published version \cite{mal2} of the e-paper \cite{mal}, the road 
to the ``well known result" recommends, possibly due to a misprint, 
the diametrically opposite prescription 
\[
\zeta = \frac{\dot{\phi}}{\dot{\rho}}f, 
\]
which would send the factor
$\gamma$ to the numerator of the above calculation. It looks as though  
the `rigorous' inflationary predictions fluctuate between 
zero and infinity.) 

In inflationary literature,
the power spectrum $P_{\cal R}(k)$ of curvature perturbations 
is usually written in the form
\[
P_{\cal R}(k) = \frac{k^3}{2 \pi^2} \frac{|u_k|^2}{z^2},
\]
where 
\[
z =a \frac{\dot{\varphi}}{H} = a {\sqrt \gamma} \sqrt{\frac{2}{\kappa}}, 
\]
and $u_k$ are the mode-functions 
($u_k = \mu_n$ in our notations) satisfying 
Eq.(\ref{dpeq}) with the initial conditions
\begin{equation}
\label{normu}
u_k = \frac{1}{\sqrt {2k}} e^{-ik \eta}~~~~~{\rm for}~~~~~
\eta \rightarrow -\infty.
\end{equation}

As one can see from the expression for $P_{\cal R}(k)$,
in inflationary theory, which is based on the initial
conditions (\ref{normu}), the divergence of $P_{\cal R}(k)$ 
in the limit of very small $\sqrt \gamma$ is present from the very 
beginning of the evolution of the perturbations. To put it differently,
the divergence takes place from the very early high-frequency regime,
where by the physical meaning of the problem we were supposed to
choose a minimal amplitude of the `gauge-invariant' metric perturbation
$\zeta$ (or, in other words, a minimal amplitude of the curvature
perturbation $\zeta$).

The crucial point of our discussion is that the temptation to 
interpret $|0_s \rangle$ as the vacuum state for the field 
$\zeta$ is, in fact, a grave error. The 
calculation of the mean-square value of the canonically 
conjugate momentum ${\bf p}$ gives
\[
\langle 0_s|{\bf p}^2|0_s \rangle =\frac{\hbar}{2} \gamma_0,
\]
so that the factor $\sqrt \gamma$ cancels out in the uncertainty
relation 
\[
\Delta {\bf q} \Delta {\bf p} =\frac{\hbar}{2}.
\]

The derived numbers clearly indicate that the 
quantum state $|0_s\rangle$ is not a genuine (ordinary) vacuum state 
$|0\rangle$ for the dynamical variable $\zeta$, but, on the
contrary, is a multi-particle (squeezed vacuum) state. This is why 
we have used the subindex $s$.

To show how the states $|0\rangle$ and $|0_s\rangle$ are related, 
we shall first transform the operators. Let us introduce the 
annihilation and creation operators ${\bf c}, {\bf c}^{\dagger}$ 
according to the Bogoliubov transformation
\begin{equation}
\label{bog}
{\bf b} = u {\bf c} + v {\bf c}^{\dagger}, ~~~~ 
{\bf b}^{\dagger} = u^{*} {\bf c}^{\dagger} + v^{*} {\bf c},
\end{equation}
where
\begin{equation}
\label{uv}
u = \cosh~r, ~~~~~v= e^{i2 \phi}~\sinh~r.
\end{equation}
The parameters $r$ and $\phi$ are called squeeze parameters.

Let us assign the following values to $r$ and $\phi$:
\begin{equation}
\label{sqp}
e^{2r} = \gamma,~~ \phi=n(\eta-\eta_0)~~~~~{\rm or}~~~~~e^{-2r} = \gamma,
~~\phi=n(\eta-\eta_0) +\frac{\pi}{2}.
\end{equation}
We shall now use the substitution (\ref{bog}), together with Eqns (\ref{uv})
and (\ref{sqp}), to Eqns (\ref{qp1d}) and (\ref{qp2d}). The factor
$1/\sqrt \gamma$ cancels out in Eq.(\ref{qp1d}) and the factor
$\sqrt \gamma$ cancels out in Eq.(\ref{qp2d}). In terms of
${\bf c}, {\bf c}^{\dagger}$, the operators ${\bf q}, {\bf p}$ will
take the final form:
\begin{equation}
\label{qp1d1}
{\bf q}= \sqrt{\frac{\hbar}{2}} \frac{\tilde{a}_0}{\tilde{a}}
\left[{\bf c}e^{-in(\eta-\eta_0)} 
+ {\bf c}^{\dagger}e^{in(\eta-\eta_0)} \right],
\end{equation}
\begin{equation}
\label{qp2d2}
{\bf p}= i\sqrt{\frac{\hbar}{2}} \frac{\tilde{a}}{\tilde{a}_0}\left[-{\bf c}
e^{-in(\eta-\eta_0)} + {\bf c}^{\dagger}e^{in(\eta-\eta_0)} \right].
\end{equation}
 
The genuine vacuum state for the variable $\zeta$ (i.e. the ground
state of the corresponding Hamiltonian) is defined by the
condition
\[
{\bf c}|0 \rangle =0.
\]
Calculating the mean square values of ${\bf q}$ and its 
canonically conjugate momentum ${\bf p}$, we find 
\[
\langle 0|{\bf q}^2|0 \rangle = \langle 0|{\bf p}^2|0 \rangle =\frac{\hbar}{2},
~~~~~~~~\Delta {\bf q} \Delta {\bf p} =\frac{\hbar}{2},
\]
as it should be.

Taking into account the definition (\ref{zbar}), we finally derive the 
initial $rms$ value of the variable $\zeta = \mu/a \sqrt\gamma$:
\begin{equation}
\label{zrmsgw}
\zeta_{rms} = {\left(\langle 0|{\bf \zeta}^2|0 \rangle\right)}^{1/2} =
\frac{\sqrt{2}(2 \pi)^{3/2} l_{Pl}}{\lambda_0}.
\end{equation}
Extrapolating the initial time $\eta_0$ up to the boundary between the
adiabatic and superadiabatic regimes at $\eta =\eta_i$, we arrive at
the estimate 
\[
\left(\frac{\mu}{a\sqrt\gamma)}\right)_{rms} \sim \frac{l_{Pl}}{\lambda_i}. 
\]
This
evaluation, plus the constancy of the quantity $\mu/a\sqrt\gamma$
throughout the long-wavelength regime, is the foundation of the result
according to which the final (at the end of the long-wavelength regime) 
amplitudes of gravitational waves and density perturbations should be
roughly equal to each other \cite{g7}. 

There is no dimensional parameter which
could be regulated in such a way as to make one of the amplitudes 
several orders of 
magnitude larger than another. In terms of the `scalar' and `tensor'
metric amplitudes this means that $h_T/h_S \approx 1$ for all $\gamma$'s. 
More accurate calculations along the same lines produce 
${\cal C} = \sqrt{24 \pi} l_{Pl}$ in expression (\ref{hij}) for 
density perturbations. 

Certainly, the correct quantisation 
procedure (\ref{qp1d1}), (\ref{qp2d2}),
as opposed to the incorrect (inflationary) procedure (\ref{qp1d}), 
(\ref{qp2d}), could be formulated from the outset of quantisation.
Mathematically, the Lagrangians (\ref{Lg2}) and (\ref{Ld2}) are alike,
if in (\ref{Lg2}) one means $\tilde{a}$ by $a$, and replaces $h$ with
$\zeta$. 

The derivation of the Hamiltonian for S-perturbations
repeats the derivation for gravitational waves. Using the canonical 
pair $q=\mu$, $p = \partial L/\partial \mu^{\prime}$ for $\mu_S$,
we arrive at the Hamiltonian (compare with Eq.(98) in Ref.\cite{g7})
\begin{equation}
\label{hamdp}
{\bf H}(\eta) = n {\bf c}^{\dagger}{\bf c} + \sigma {{\bf c}^{\dagger}}^2 +
\sigma^{*} {\bf c}^2,
\end{equation}
where the coupling to the external field is now given by the function
$\sigma(\eta) = (i/2)(\tilde{a}^{\prime}/{\tilde{a}})$. 

The Heisenberg equations of motion for the Heisenberg operators
${\bf c}(\eta)$, ${\bf c}^{\dagger}(\eta)$ lead to 
classical equation (\ref{dpeq}). The asymptotic expressions for the
Heisenberg operators,
\[
{\bf c}(\eta) = {\bf c} e^{-in(\eta- \eta_0)}, ~~~~~
{\bf c}^{\dagger}(\eta) = {\bf c}^{\dagger} e^{in(\eta-\eta_0)},
\] 
are participating in Eqs. (\ref{qp1d1}), (\ref{qp2d2}). Clearly, the 
vacuum state $|0\rangle$, defined as ${\bf c}(\eta)|0\rangle =0$, 
minimises the oscillator's energy (\ref{hamdp}). 

Since at times 
near $\eta_0$ the coefficients $a/a_0$ and $\tilde{a}/\tilde{a}_0$
are close to 1, the equality of the initial values for $h_{rms}$ and
$\zeta_{rms}$ follows already from the simple comparison of the
Lagrangians (\ref{Lg2}) and (\ref{Ld2}).

The relationship between the above-mentioned genuine vacuum state 
$|0\rangle$ and the squeezed vacuum state $|0_s\rangle$ is 
determined by the action of the squeeze operator $S(r, \phi)$
on $|0\rangle$:
\[
|0_s\rangle = S(r, \phi) |0 \rangle ,
\]
where
\[
S(r, \phi) = \exp\left[\frac{1}{2}r \left(e^{-i2\phi}{\bf c}^2 -
e^{i2\phi} {{\bf c}^{\dagger}}^2 \right) \right].
\]
The mean number of quanta in the squeezed vacuum state is given by
\[
\langle 0_s|{\bf c}^{\dagger}{\bf c}|0_s \rangle = 
\sinh^2 r=\frac{1-\gamma}{2 \sqrt \gamma}.
\]

This is a huge and divergent number, when the `zero in the denominator'
factor $\sqrt \gamma$ goes to zero. Therefore, the ``standard 
inflationary result" for S-perturbations is based on the wrong 
initial conditions, according to which the initial amplitude of the 
$\zeta$-perturbations can be arbitrarily large from the very beginning 
of their evolution. 

Moreover, the initial amplitude is assumed to go to 
infinity in the most interesting limit of $\sqrt \gamma \rightarrow 0$ 
and ${\rm n} \rightarrow 1$. If $\sqrt \gamma$ does not deviate from
1 too much, then the mean number of quanta in the squeezed
vacuum state is acceptably small, and the wrong initial conditions
give results sufficiently close to the correct ones. However, as in the
Landau example mentioned above, if the wrong formula gives acceptable
answers for some range of $x$, this does not make the wrong theory
a correct one. (Finally, if $\sqrt \gamma =1$, then $a(t) \propto t$, $a(\eta)
\propto e^{\eta}$, $w= -1/3$. From this model of cosmological
evolution the study of relic gravitational waves has began in the 
first paper of Ref.\cite{g1}.) 

In terms of the classical mode functions,
it is the function $\mu/a \sqrt \gamma$ that should satisfy the
classical version of the initial
conditions (\ref{qp1d1}), and not the function $\mu/a$, which is postulated
by the inflationary requirement (\ref{normu}). 
They both are so-called `gauge-invariant' variables, but their 
physical meaning is drastically different. The original derivations 
of the ``well-known result" were guided simply by the
visual analogy between the function $u=\mu$ in theory of density 
perturbations and the function $\mu$ in the
theory of gravitational waves already developed by that time. 

The assumption of 
arbitrarily large initial amplitudes of curvature perturbations or, 
technically speaking, the choice of the initial multi-particle 
squeezed vacuum state $|0_s\rangle$ for $\zeta$, instead 
of the ordinary vacuum sate $|0 \rangle$, is
the origin of the absurd ``standard inflationary result". Certainly, 
this wrong assumption cannot be the basis of observational predictions 
for cosmology.   

\section{Conclusions} 

The grossly incorrect predictions of inflationary theorists should not
be the reason for doubts about the existence and expected amount of relic
gravitational waves. The generation of relic gravitational waves is
based on the validity of general relativity and quantum 
mechanics in a safe cosmological regime where quantisation of the
background gravitational field is not necessary. 

In our numerical
evaluations, we also assumed that the observed large-angular-scale 
anisotropies of CMB are caused by cosmological perturbations of 
quantum-mechanical origin. This is not necessarily true, but it would 
be quite disastrous if it proved to be untrue. 

It is quite a challenge
to imagine that the natural and unavoidable quantum-mechanical generation
of cosmological perturbations is less effective than anything else. 
In any case, if relic gravitational waves are not discovered at 
the (relatively high) level described in this contribution, the 
implications will be much more serious than the rejection of one
inflationary model or another. The reality of our time is such that if 
the proposal is not properly `sexed-up', it is not very likely to be 
funded. But the ultimate truth lies in the fact that the real physics 
of the very early Universe is much more exciting than the 
artificial hullaballoo over
popular words such as `inflation' or `inflationary gravitational waves'.

Hopefully, relic gravitational waves will be discovered in experiments,
which are already in the well-developed stage. I personally would
think that this is likely to happen first in dedicated ground-based
observations, such as the recently approved Cardiff-Cambridge-Oxford 
collaboration CLOVER \cite{taylor}. Let us hope this will indeed be the case.

\section*{Acknowledgments} 

I am grateful to D. Baskaran, J. Romano and especially M. Mensky 
for fruitfull discussions and help, and to P. Steinhardt for calling 
my attention to the paper \cite{mal2} and for the accompanying 
intense and useful correspondence.


\end{document}